# On the Response of Basic Walfisch-Ikegami and Walfisch-Bertoni Models to QMM Calibration

Ayotunde Ayorinde
Department of Electrical & Electronics Engineering
University of Lagos, Akoka, Lagos, Nigeria

Hisham Muhammed
Department of Electrical & Electronics Engineering
University of Lagos, Akoka, Lagos, Nigeria

A. Ike Mowete
Department of Electrical & Electronics Engineering
University of Lagos, Akoka, Lagos, Nigeria

## ABSTRACT
This paper systematically examines the response to Quasi-Moment-Method (QMM) calibration, of the basic COST231-Walfisch-Ikegami, ITUR-Walfisch-Ikegami, and Walfisch-Bertoni models. First, it is demonstrated that the component parameters of the models are suitable candidates for use as expansion/testing functions with QMM pathloss model calibration schemes; and thereafter, the basic models are subjected to calibration, using measurement data available in the open literature. Computational results reveal that the COST231-Walfisch-Ikegami and ITU-Walfisch-Ikegami models have virtually identical QMM-calibration root mean square error (RMSE) responses; and that the Walfisch-Bertoni model has better RMSE responses than both of them. A particular attribute revealed by the simulation results is that all QMM-calibrated 'Walfisch-type' basic models have excellent mean prediction error (MPE) metrics (in general, less than 0.001dB). In addition to pathloss prediction profiles, the paper also presents profiles of disaggregated net pathloss, in terms of contributions by component parameters, including 'roof-top-to-street' diffraction and scatter loss and multiscreen diffraction loss.

## 1. INTRODUCTION
Models developed and utilized for the predictions of energy lost along the propagation path, by propagation radiowaves are generally classifiable as either empirical (or statistical) or site-specific (deterministic/semi-deterministic) models, [1]. Of the existing site-specific models, the Walfisch-Bertoni [2] and the Walfisch-Ikegami models [3] are arguably the best known, and have, in recent times, been the subject of quite a few studies, [4]-[11]. Although a significant number of the investigations (for example, [9], [10]) merely evaluated the performance of the basic model in various propagation environments and scenarios, many others involved statistical 'tuning' of the basic model (as in [11], [13]), least-square regression optimization [7], and meta-heuristics-based optimization, [4], for the purposes of improving model prediction accuracy. In the cases concerning evaluation limited only to the basic Walfisch-Ikegami models, outcomes typically indicate that some other basic model (usually the COST231-Hata model) provides better prediction metrics, and consequently selected for further processing, in most cases, least-square-regression, for performance optimization. As example, Mohammed and Jaafer, [10], reported that the basic COST231-Walfisch Ikegami model gave an average (0ver four sites) prediction RMSE of 12.4dB compared with a corresponding value of 10.8dB by the COST231-Hata model, to inform the choice of the latter for least square regression optimization. In a similar contribution concerning two LTE networks, Imoize et al, [15] reported simulation results, one set for a sub-urban environment, and for which the basic Ericsson model with a prediction RMSE of 7.0797dB compared to 13.8738dB recorded for the corresponding COST231-Walfisch-Ikegaimi model, was selected for optimization. The other set of results reported in [15] for a metropolitan area, had the basic COST231-Hata model recording 5.1343dB (as against 5.2496 dB for the COST231-Walfisch-Ikegami) to emerge as candidate for least square regression optimization. Omer et al. [7], also presented results, which, in terms of 'pathloss exponents' (3.7191 and 3.8004 for the Hata and Walfisch-Ikegami models, respectively) suggested that for large urban environments, and at 'LTE frequencies, performances of the basic COST231-Hata and Walfisch-Ikegami models are comparable.. On the other hand, some other investigations focused on 'tuning' the basic COST231-Walfisch-Ikegami, for improved prediction accuracy. Examples of these include the contribution by Alqudah, [8], whose use of a MATLAB curve fitting routine resulted in an improvement of RMSE from 6.78dB for the basic model to 4.46dB, in a Non-Line-of-Sight (NLOS) case: and from 22.6dB to 6.38dB for a Line-of-Sight (LOS) example. 'Statistical tuning' of the basic COST231-Walfisch-Ikegami model is one approach that has received the attention of many researchers. Ambawade et al. [6] reported a statistical tuning approach based on multiple regression, in which building height, building separation, and street orientation component parameters of the basic model, were prescribed as Gaussian random variables. RMSE metrics recorded for the tuned models ranged between 3.1354dB and 8.1373dB. A similar approach was adopted by Rozal and Pelaes, [13], whose model assigned Gaussian random variable distributions to building height and building separation. Although the paper did not provide RMSE metrics for the tuned model, the 'average error' values of between 0.0195dB and 4.1825dB reported in the paper, as well as the profiles displayed in the associated graphical illustrations suggest that the tuned model provided very good prediction results. In addition to a statistical tuning technique involving multiple linear regression with building height, street width, and street orientation specified as Gaussian random variables, Tahat and Taha, [11], also utilized a Particle Swarm Optimization (PSO) algorithm towards improving the performance of the basic model. Bhuvaneshwari et al. [4], presented the use of three different meta-heuristic algorithms (Genetic Algorithm (GA), PSO, and Grey-Wolf Optimization (GWO)) for tuning the basic COST-231 Walfisch-Ikegami prediction model. The optimization processes each had four model parameters ('distance', 'transmitter antenna-building heights', 'building separation', and 'street orientation angle') as the candidates for optimization, and three parameters were prescribed by the authors as 'normalized random variables', characterized by Gaussian probability distributions. Optimization was effected with the use of 'real time' (instantaneous) measurements of received power from a deployed 900MHz transmitter, and RMSE metrics were recorded as 0.4251dB, 0.2939dB, and 0.2248dB for the GA, PSO, and GWO optimized models, respectively.

This paper investigates the response of five basic (nominal) 'Walfisch-type' models to calibration with the use of the Quasi-Moment-Method (QMM) technique, described in a recent publication by Adelabu et al. [19]. The basic models are the (NLOS) COST231-Walfisch-Ikegami for Metropolitan centers and sub-urban regions [3]; a variant of these models including a correction pointed out by Jeong and Lee [5], as implemented by ITU-R [16], and the Walfisch-Bertoni model, [2]. Measurement data utilized for calibration were obtained from the open literature, [4], [10], and [15], using the commercial graph digitizer software, 'getdata'. A number of interesting features emerged from the simulation results: first, the two QMM-calibrated COST231 (metropolitan and sub-urban) and their ITU-R counterparts, with only a few exceptions, recorded identical RMSE metrics. Second, all the QMM-calibrated 'Walfisch-type' models recorded remarkably low mean prediction error (MPE) metrics, suggesting that they are particularly suitable for cases where MPE is the metric of choice. And third, the QMM-calibrated Walfisch-Bertoni models recorded better metrics than the Walfisch-Ikegami models, matching (and even surpassing, in some cases) the metrics due to the ECC33 models, utilized as benchmark. The computational results also suggest that choice of basic model as candidate for linear regression optimization need not be be informed by the relative RMSE performance of the basic models.

## 2. THEORY

As noted by Harrington [17], one very important consideration for the successful implementation of Method of Moments (and hence, QMM) schemes, concerns choice of expansion and testing functions. In particular, [17], expansion functions should be linearly independent, and should be such that some superposition of the functions provides a reasonable approximation to the desired function. Testing functions are also required to be linearly independent, and be such that the evaluation of the inner product quantities involving the testing function and field measurement data is essentially determined by the relatively independent properties of the field measurement data. And according to theorem 4.2.5 of [18], linear independence of the expansion and testing functions guarantees uniqueness of the solution. Because the basic models evidently satisfy the requirement of providing 'reasonable approximations' it only remains to establish linear independence for the basic models of interest to this paper.

In that connection, the COST231 Walfisch-Ikegami models of interest here are first identified as given by [3], [17],

$$P_{loss} = L_{fsp} + L_{rts} + L_{msd}, \quad (1)$$

where the free-space loss denoted by $L_{fsp}$ admits description according to

$$L_{fsp} = 32.4 + 20\log_{10}d + 20\log_{10}f \quad (1a)$$

and the roof-top-to-street diffraction and scatter loss, $L_{rts}$, for the cases of interest is either given by

$$L_{rts} = -16.9 - 10\log_{10}w + 10\log_{10}f + 20\log_{10}\Delta h_{rx} + 0.354\varphi - 10, \quad (1c)$$

when $0 \leq \varphi \leq 35^0$: or

$$L_{rts} = -16.9 - 10\log_{10}w + 10\log_{10}f + 20\log_{10}\Delta h_{rx} + 2.5 + 0.075(\varphi - 35^0), \quad (1d)$$

for $35^0 \leq \varphi \leq 55^0$. A third possibility for $L_{rts}$ owes to a correction pointed out by Jeong and Lee, [5], and implemented in the ITUR-R model, such that the leading terms of the right hand members of Eqn. (1c) and 1(d) become -8.2 rather than -16.9. The multiscreen diffraction loss, $L_{msd}$, is also, for this paper's purposes, described by two possible expressions; the first, applicable for metropolitan centers, is

$$L_{msd} = -18\log_{10}(1+\Delta h_{tx}) + 54 + 18\log_{10}d + \left(1.5\left(f/925 - 1\right) - 4\right)\log_{10}f - 9\log_{10}b, \quad (1e)$$

whilst in the case of suburban cities,

$$L_{msd} = -18\log_{10}(1+\Delta h_{tx}) + 54 + 18\log_{10}d + \left(0.7\left(f/925 - 1\right) - 4\right)\log_{10}f - 9\log_{10}b \quad (1f)$$

It should also be noted that for eqn. (1f), when operating frequency is greater than 2GHz, the second term on the right hand side (54) is replaced by 71.4, [16]. In Eqns. (1), 'd' in km, stands for radial distance from transmitting antenna, 'f', operating frequency in MHz, '$\Delta h_{rx}$', (m), difference between roof top height and mobile station height, '$\Delta h_{tx}$', (m) difference between transmitter antenna height and roof top height; 'w'(m) represents street width, whilst 'b' (m) stands for the separation of buildings. The symbol '$\varphi$' appearing in Eqns. (1c) and (1d) represents 'the angle of incidence', in degrees, of the propagating wave with respect to the location of measurements, and defines 'street orientation'.

Finally, the basic Walfisch-Bertoni model is given [2], by

$$P_{loss} = L_{fsp} + L_{exs}, \quad (2)$$

for which the 'free space' loss remains as defined by Eqn. (1b), and the 'excess pathloss' is given by

$$L_{exs} = 57.1 + \log_{10}f + 18\log_{10}d - 18\log_{10}\Delta h_{tx} - 18\log_{10}\left(1 - \frac{d^2}{17\Delta h_{tx}}\right) + A, \quad (2a)$$

provided that 'A', which in [2], is described as accounting for the influence of the building geometry, is given by

$$A = 5\log_{10}\left[\left(b/2\right)^2 + \left(\Delta h_{rx}\right)^2\right] - 9\log_{10}b + 20\log_{10}\left[\tan^{-1}\left(2\Delta h_{rx}/b\right)\right]. \quad (2b)$$

All the terms in Eqns. (2a) and 2(b) remain as earlier defined, with the argument af the arc-tangent function specified in degrees. The proposal in this paper is to utilize the component parameters of the pathloss expressions of Eqns. (1) and (2) as expansion and testing functions in QMM-model-calibrations.

And as earlier explained, one fundamental requirement is to establish that the set prescribed in the case of the Walfisch-Ikegami models, as represented, for example, by

$$\{f_n\}_{n=0}^{12} = \begin{Bmatrix} 32.4, 20\log_{10}d, 20\log_{10}f, -16.9, -10\log_{10}w, 10\log_{10}f, \\ 20\log_{10}\Delta h_{rx}, (-10+0.354\varphi), -18\log_{10}(1+\Delta h_{tx}), 54, \\ 18\log_{10}d, \left(1.5\left(f/925 - 1\right)\right)\log_{10}j, -9\log_{10}b \end{Bmatrix}$$
(3)

represent a set of linearly independent functions. Now, for a given transmitter, mobile station, and terrain, any linear combination of these functions will assume the form

$$g_1(d) = c_1 + c_2 \log_{10} d,  \quad (4)$$

whose Wronskian is readily evaluated as

$$W(g_1) = \det\left(\begin{bmatrix} c_1 & c_2 \log_{10} d \\ 0 & c_2/d\log_e 10 \end{bmatrix}\right) = c_1 c_2 / d\log_e 10. \quad (5)$$

Because $W(g_1)$ is non-zero for all finite values of 'd', ($c_1$ and $c_2$ being constants) the functions satisfy the requirement of linear independence, [20]. In the case of the basic Walfisch-Bertoni model, the indicated expansion (and testing) functions are

$$\{f_n\}_{n=0}^7 = \begin{Bmatrix} 89.5, 38\log_{10}d, -18\log_{10}\Delta h_{tx}, 21\log_{10}f, \\ 5\log_{10}\left[(b/2)^2 + (\Delta h_{rx})^2\right], -9\log_{10}b, \\ 20\log_{10}\left[\tan^{-1}\left(2\Delta h_{rx}/b\right)\right], -18\log_{10}\left(1 - \dfrac{d^2}{17\Delta h_{tx}}\right) \end{Bmatrix}$$
(6)

Any linear combination of the functions in Eqn. (6) will be of the form

$$g_2(d) = k_1 + k_2 \log_{10} d + k_3 \log_{10}(1 - k_4 d^2), \quad (7)$$

and it is a straightforward matter to show that

$$W(g_2) = \det \begin{bmatrix} k_1 & k_2 \log_{10} d & k_3 \log_{10}(1 - k_4 d^2) \\ 0 & \dfrac{k_2}{d\log_e 10} & \dfrac{k_3}{\log_e 10}\left(\dfrac{-2k_4 d}{1 - k_4 d^2}\right) \\ 0 & \dfrac{-k_2}{d^2 \log_e 10} & \dfrac{k_3}{\log_e 10}\left(\dfrac{2k_4(1 + k_4 d^2)}{d^2(1 - k_4 d^2)}\right) \end{bmatrix} \quad (8)$$

$$= \dfrac{4k_1 k_2 k_3 k_4^2 d}{(1 - k_4 d^2)(\log_e 10)^2}, \quad k_4 d^2 \neq 1$$

Again, it is seen that the functions prescribed by Eqn. (6) are according to Eqn. (8), linearly independent. With these functions available for use therefore, the QMM-calibration process reduces to the solution of the approximation problems defined by

$$P_{L-QMM-WI}(d) = \alpha_0 f_0 + \alpha_1 f_1 + \ldots + \alpha_{12} f_{12} = P_{mea}(d), \quad (9)$$

for the Walfisch-Ikegami models, and

$$P_{L-QMM-WB}(d) = \alpha_0 f_0 + \alpha_1 f_1 + \ldots + \alpha_7 f_7 = P_{mea}(d), \quad (10)$$

for the Walfisch-Bertoni models. The solutions to the problems emerge, when, in each case, the desired unknown calibration coefficients are, as explained in [19], obtained as

$$(\{\alpha_n\}_{N=1}^N)^T = \begin{bmatrix} \langle f_0, f_0 \rangle & \langle f_0, f_1 \rangle & \cdots & \langle f_0, f_N \rangle \\ \langle f_1, f_0 \rangle & \langle f_1, f_1 \rangle & \cdots & \langle f_1, f_N \rangle \\ \cdots & \cdots & \cdots & \cdots \\ \langle f_N, f_0 \rangle & \langle f_N, f_1 \rangle & \cdots & \langle f_N, f_N \rangle \end{bmatrix}^{-1} \begin{pmatrix} \langle f_0, P_{mea} \rangle \\ \langle f_1, P_{mea} \rangle \\ \cdots \\ \langle f_N, P_{mea} \rangle \end{pmatrix}$$
(11)

## 3. COMPUTATONAL RESULTS
### 3.1 Calibration with Measurements from [10]

Using measurements and associated terrain parameters available from [10] for a 900MHz deployed network in Karada district of Baghdad, all the five models described in §2 were subjected to QMM-calibration. As can be seen from Table I of [10], three of the four 'sectors' considered in [10] share identical terrain parameters, though profiles of measured pathloss differ.

Representative examples of outcomes of the calibration process are displayed in Fig. (1) for 'sector 1', and Fig. (2), for sector 2. The illustrations also include pathloss profiles for ECC33 (large and medium cities) calibrated as described in [19]; with the same sets of measurements and utilized here as benchmark, on account of established excellent response to QMM-calibration, [19]. It is apparent from the illustrations that the Walfisch-Ikegami models are characterized by good responses to QMM-calibration, and that the response of the Walfisch-Bertoni model is significantly better than those of the other Walfisch-type models.

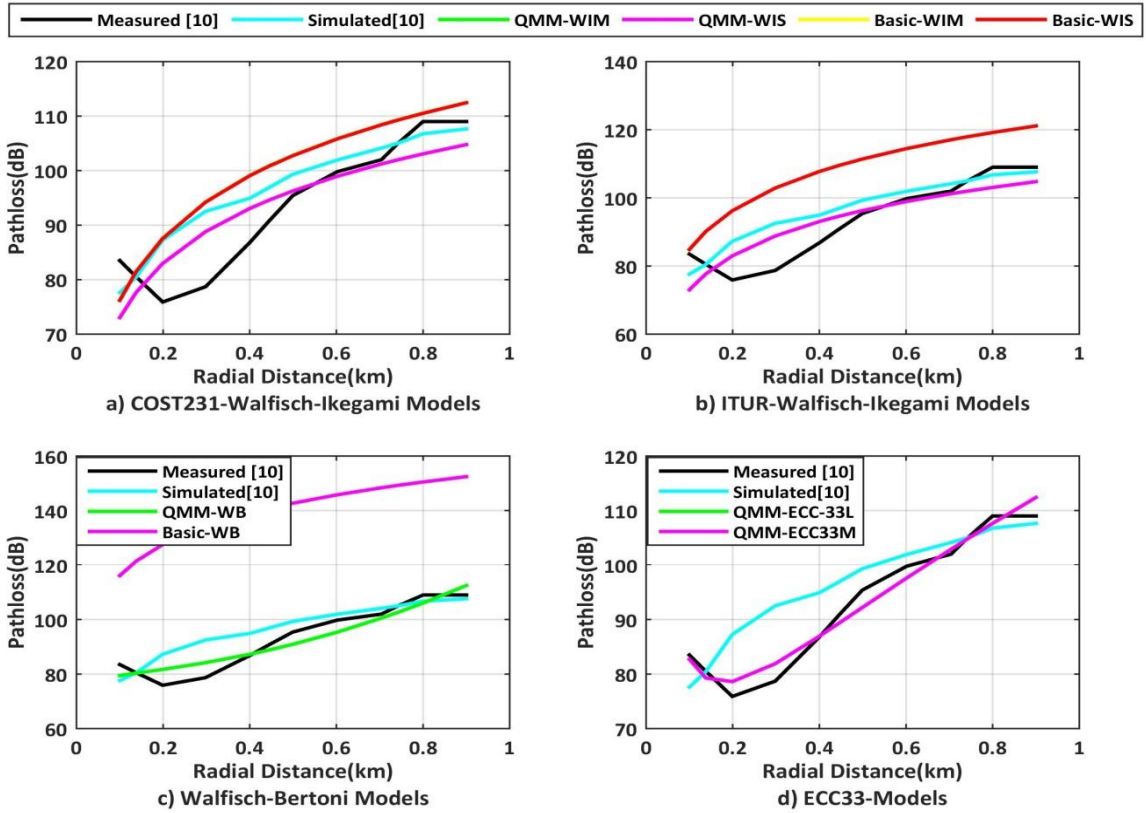

**Fig 1: Profiles of pathloss predicted by Models QMM-calibrated with measurement data for sector 1 of [10]**

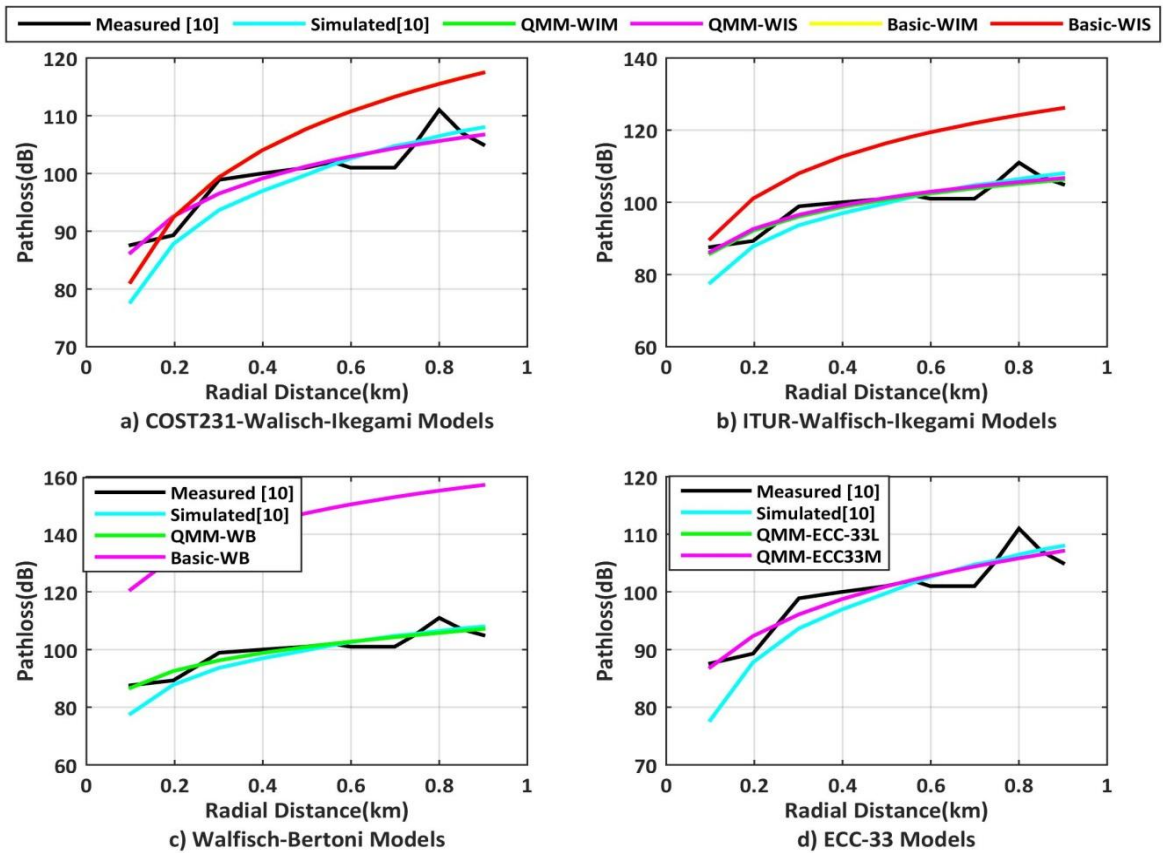

**Fig 2: Profiles of pathloss predicted by Models QMM-calibrated with measurement data for sector 2 of [10]**

**Table 1. RMSE and MPE Metrics for the Pathloss Profiles of Fig. 1**

| Calibrated Models | Sector1 | | Sector2 | | Sectir3 | | Sector4 | |
|---|---|---|---|---|---|---|---|---|
| | RMSE | MPE | RMSE | MPE | RMSE | MPR | RMSE | MPE |
| CWI-M | 5.6818 | -0.0043 | 2.3558 | 0.0020 | 4.8261 | 0.0023 | 5.5087 | -0.0071 |
| CWI-SU | 5.6818 | 0.0011 | 2.3558 | 0.0003 | 4.8261 | 0.0011 | 5.5087 | 0.0027 |
| ITWI-M | 5.6818 | 0.0005 | 2.4036 | 0.4768 | 4.8261 | 0.0014 | 5.5087 | 0.0008 |
| ITWI-SU | 5.6818 | 0.0003 | 2.3558 | 0.0040 | 4.8261 | 0.0009 | 5.5087 | 0.0004 |
| W-BERT | 3.5611 | 0.5441 | 2.3394 | 0.0018 | 4.6644 | 0.0029 | 3.4090 | 0.0002 |
| ECC33-L | 2.0613 | 0.0050 | 2.3337 | 0.0041 | 4.8241 | 0.0006 | 2.4467 | 1.3368 |
| ECC33-M | 2.0613 | 0.0000 | 2.3338 | 0.0044 | 4.8241 | 0.0014 | 2.0388 | 0.0021 |

These observations are supported by the statistical metrics displayed in Table 1 for all four sectors considered in this case.

The metrics in Table 1 reveal that with the exception of one instance, all the Walfisch-Ikegami models (COST231-Metropolitan and Suburban (CWI-M/CWI-SU) as well as the corresponding ITU-R models (ITWI-M/ITWI-SU) have identical RMSE responses, ranging between 2.3558dB (for sector2) and 5.6818dB, for sector 1. As good as these metrics are, the MPE metrics are remarkably better, as can be seen from the table. The worst value of MPE = 0.0071dB recorded by CWI-M, for sector 4 is clearly an excellent outcome.

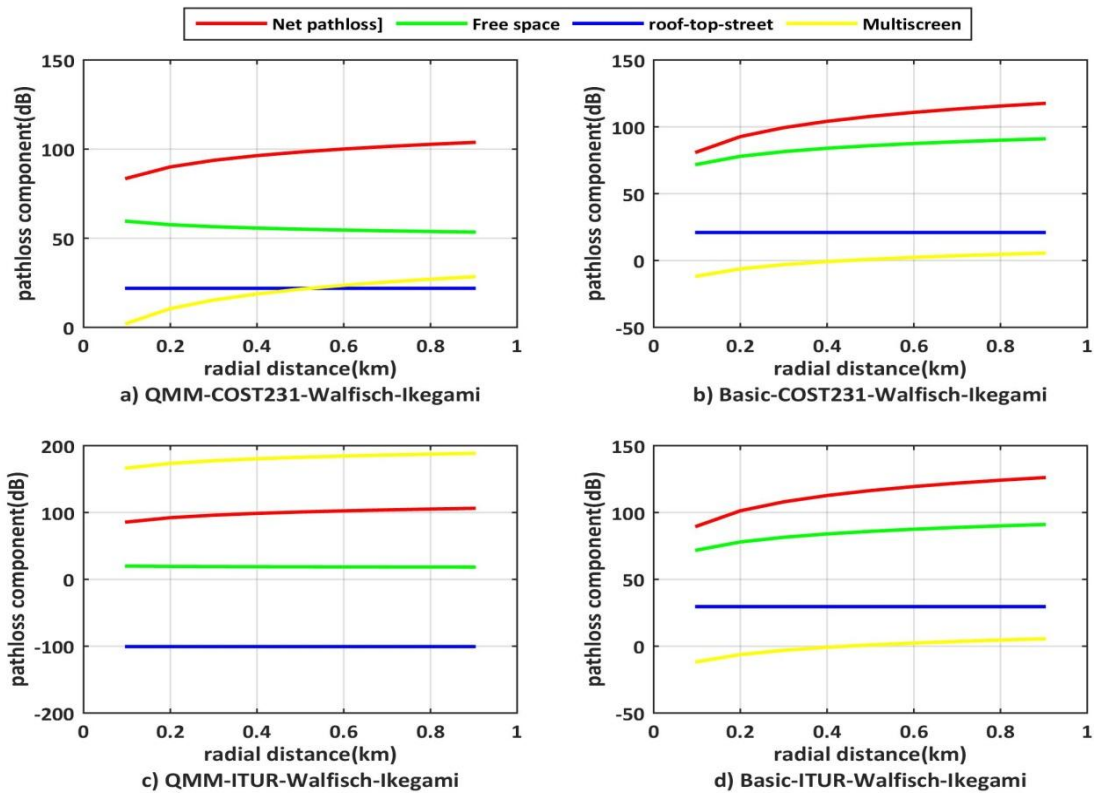

**Fig 3: Contributions to net predicted pathloss by component parameters of basic and QMM –calibrated Walfisch-Ikegami models for sector 2 of [10]**

The QMM-calibrated Walfisch Bertoni models can be seen to have better RMSE metrics (comparable with those for the ECC33 models) than the corresponding Walfisch-Ikegami models; though the MPE performances are in general, comparable, for both model types. Computational results show that the RMSE metrics represent significant improvements over corresponding values for the basic models. In the case of sector 1, for example, the percentages improvements are 32.63, 32.29, 64.04, 62.34, and 92.26 for the COST231-Walfisch Ikegami (Metropolitan and suburban), ITU-R-Walfisch-Ikegami (Metropolitan and suburban) and Walfisch-Bertoni models, respectively. The illustrations of Fig. (3) describe the profiles of the net pathloss predicted by the calibrated and corresponding basic Walfisch-Ikegami models, when disaggregated into its component parts, in the case of 'sector 2', as an illustrative example.. It is clear from the profiles that whereas the QMM-calibrated COST231 and ITU-R models, on the aggregate, have virtually identical RMSE responses, the responses of their component parameters (particularly multiscreen diffraction) to QMM-calibration differ considerably.

### 3.2 Calibration with Measurements from [4]

All five 'Walfisch-type' models were QMM-calibrated with measurements available from [4], also concerning a 900MHz network, but in which the field measurements are real-time (instantaneous) pathloss quantities, rather than averaged values.

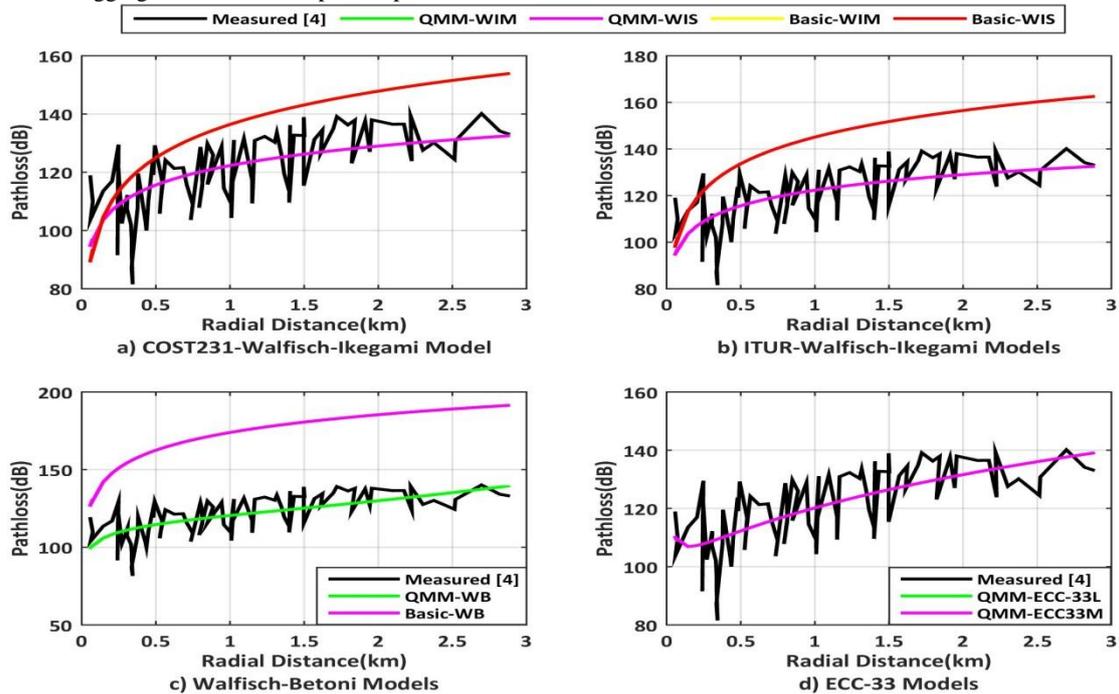

**Fig 4: Profiles of pathloss predicted by Models QMM-calibrated with measurement data from [4]**

**Table 2. RMSE and MPE Metrics for the Pathloss Profiles of Fig. 3**

|  | QMM-CALIBRATED |  | BASIC |  |
|---|---|---|---|---|
| **Model** | **RMSE** | **MPE** | **RMSE** | **MPE** |
| CWI-M | 10.3246 | 0.0056 | 17.5622 | -12.8903 |
| CWI-SU | 10.3246 | -0.0023 | 17.5197 | -12.8324 |
| ITWI-M | 10.3246 | -0.0006 | 24.5785 | -21.4903 |
| ITWI-SU | 10.3246 | -0.0013 | 24.5891 | -21.5024 |
| W-BERT | 10.1082 | -0.0006 | 51.7160 | -50.3217 |
| ECC33-L | 9.6150 | -0.0032 | N/A | N/A |
| ECC33-M | 9.6150 | 0.0001 | N/A | N/A |

It is apparent from the profiles of Fig. (4) that the QMM-calibrated Walfisch-Bertoni models have better responses than the corresponding Walfisch-Ikegami (COST-231 and ITU-R, metropolitan and suburban) models. And the statistical performance metrics of Table (2) confirm this observation.

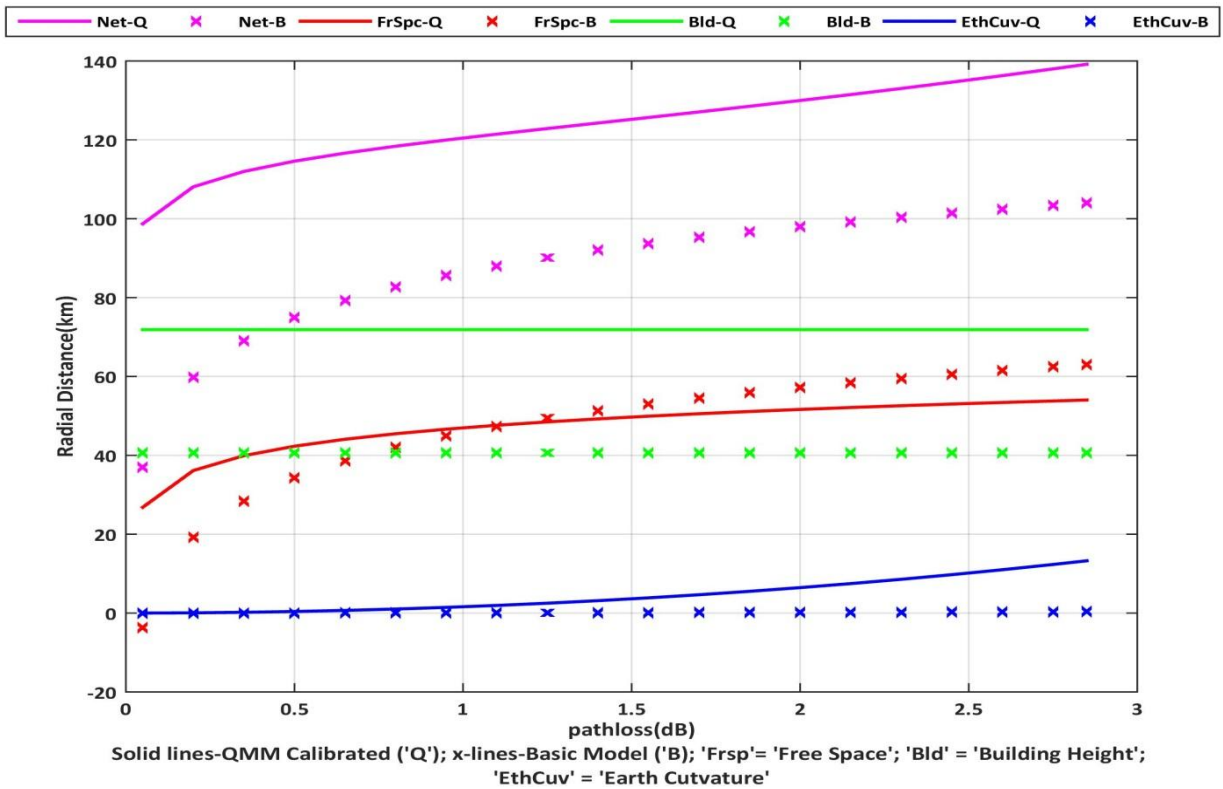

**Fig 5: Contributions to net predicted pathloss by component parameters of basic and QMM–calibrated Walfisch-Bertoni models**

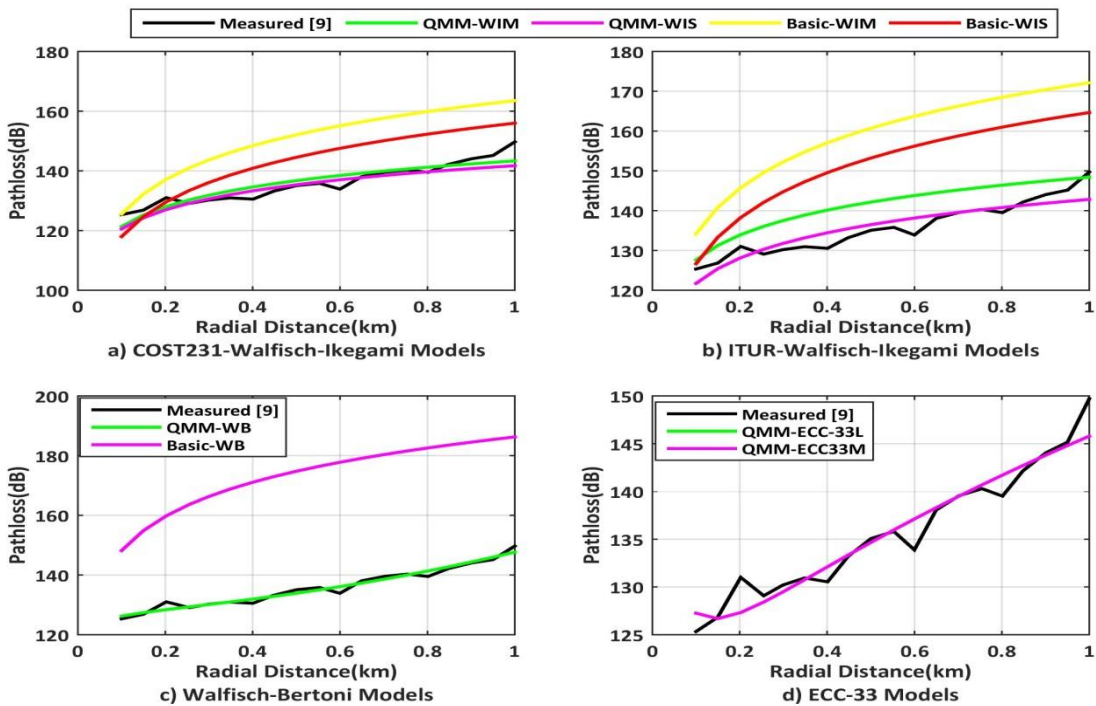

**Fig 6: Profiles of pathloss predicted by the Walfisch-type models, calibrated with measurements for the metropolitan area of [15]**

Again, the four Walfisch-Ikegami models recorded identical RMSE metrics of 10.3246, which for the QMM-COST231 metropolitan model, represents a 41.11% improvement over that for the corresponding basic model. Percentage improvements for the other QMM-calibrated models (COST231-subsurban; ITU-R-metropolitan, ITU-R-suburban, and Walfisch-Bertoni) are readily verified as being 40.98%, 57.9%, 58%, and 80.45%, respectively. Although the RMSE metrics in excess of 10dB may ordinarily be regarded as not particularly satisfactory, the fact that the field measurements with which calibration was effected are 'real-time' quantities characterized by sharp spikes suggest that the RMSE results should be considered very good.

**Table 3. RMSE and MPE Metrics for the Pathloss Profiles of the basic models calibrated with measurements from [15]**

| Model | Metropolitan | | Sub-Urban | |
|---|---|---|---|---|
| | RMSE | MPE | RMSE | MPE |
| CWI-M | 2.7142 | -0.2018 | 5.4478 | 0.6260 |
| CWI-SU | 2.9219 | 1.1277 | 5.4047 | -0.0016 |
| ITWI-M | 6.2886 | 5.6810 | 5.4047 | -0.0023 |
| ITWI-SU | 2.6954 | 0.0007 | 5.4047 | 0.0029 |
| W-BERT | 1.1941 | 0.0018 | 2.6036 | 0.0025 |
| ECC33-L | 1.6598 | 0.0021 | 3.0728 | -0.0043 |
| ECC33-M | 1.6598 | 0.0018 | 3.0727 | 0.0019 |

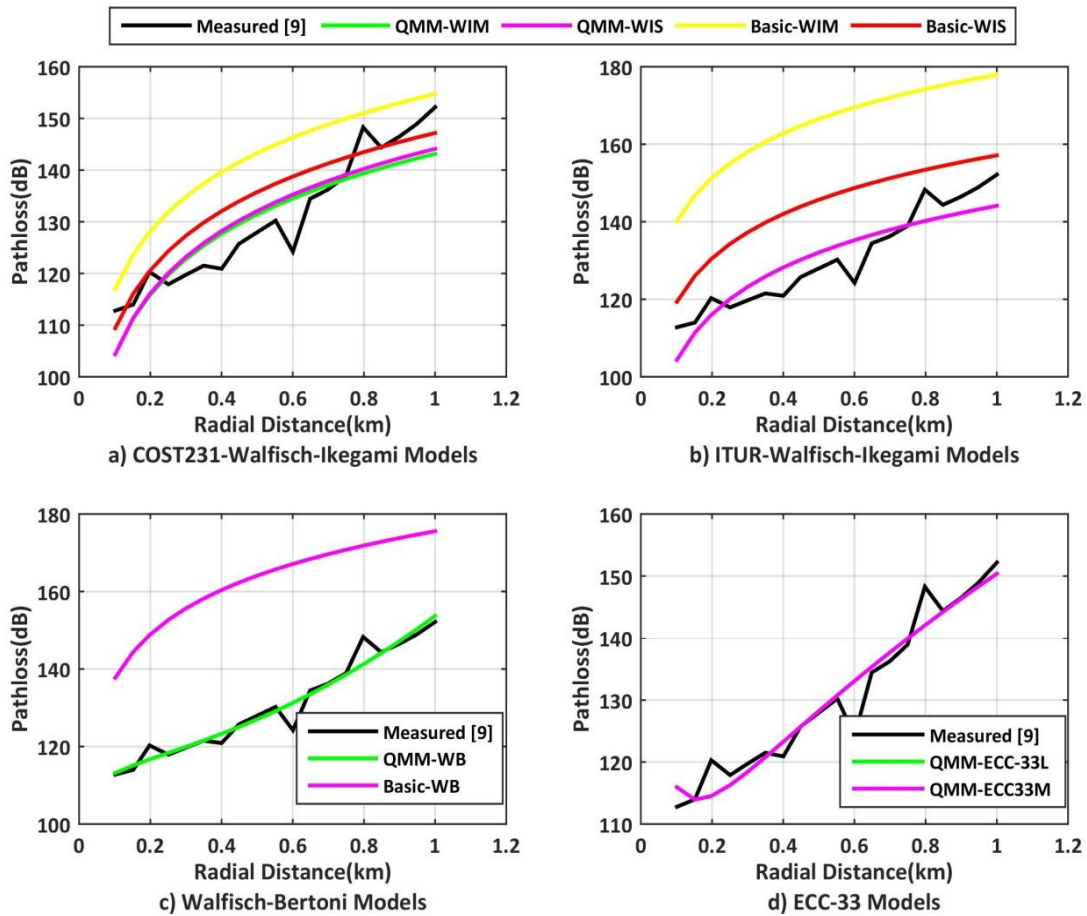

**Fig 7: Profiles of pathloss predicted by the Walfisch-type models, calibrated with measurements for the suburban area of [15]**

Indeed, the excellent MPE values recorded in all cases very strongly support this foregoing observation.

Figure (5) displays the distributions of disaggregated net-pathloss due to component parameters of the QMM-calibrated and basic Walfisch-Bertoni models concerning Fig. (4). A comparison of the contributions due to the basic model and the QMM-calibrated model shows that the biggest effect of QMM-calibration are manifested in the 'influence of the building geometry', [2]. The profiles of Fig. (5) reveal that close to the transmitter, and up to about 1.5km away from it, contributions due to the 'effects of the curvature of the Earth' [2], are about the same for the basic and QMM-calibrated models; thereafter, the QMM-calibrated model increasingly contributes significantly more than the basic model. The free-space component of the basic model contributes less to its corresponding net pathloss than does that of the QMM-calibrated model, for distances less than 1km away from the transmitter. Further away than this, its contributions to the net pathloss becomes larger, compared with those by the QMM-calibrated model.

### 3.3 Calibration with Measurements from [15]

The QMM-calibrations in this case, utilized field measurements reported by Imoize et al. [15], for a deployed LTE network, operating at 3.4GHz. Measurements were taken at two sites, one described [15], as 'metropolitan', and the other, 'suburban', with site parameters of interest to the QMM-calibration of interest here, displayed in Table (1) of [15].

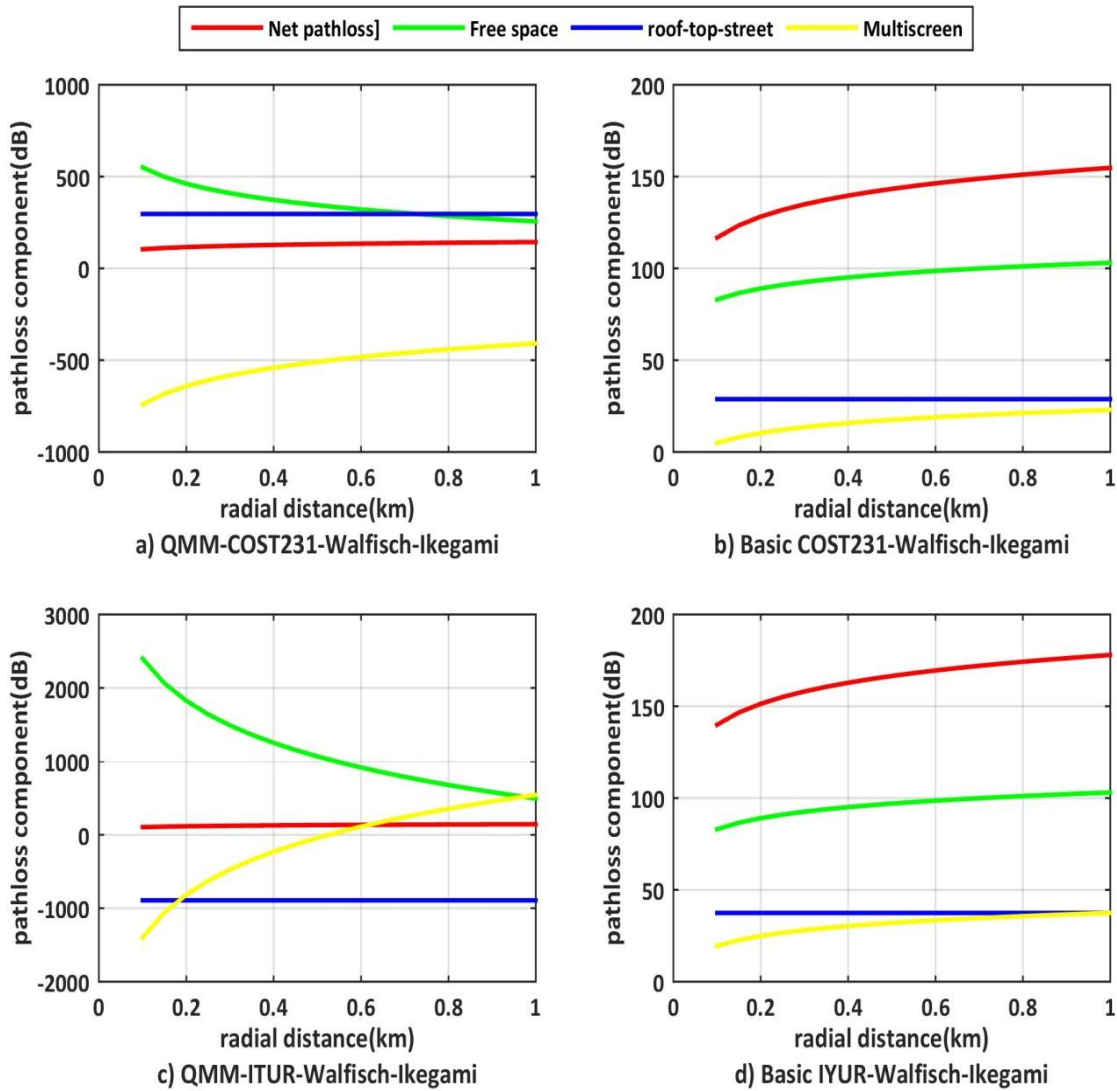

**Fig 8: Profiles of contributions to net pathloss predicted by the Walfisch-Ikegami models, calibrated with measurements for the metropolitan area of [15]**

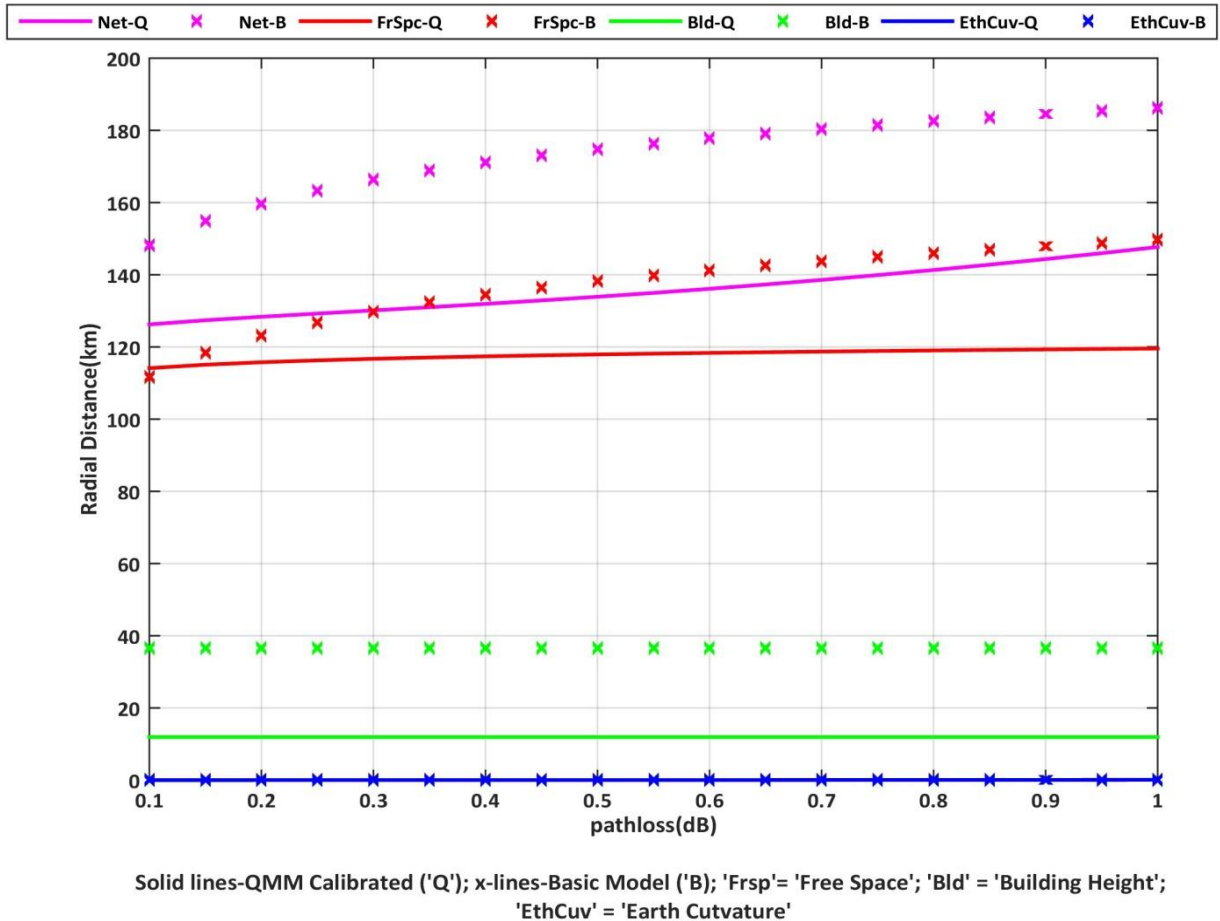

**Fig 9: Profiles of contributions to net pathloss predicted by the Walfisch-Bertoni model, calibrated with measurements for the metropolitan area of [15]**

Statistical performance metrics of RMSE and MPE corresponding to the profiles of figure (6) are shown in Table 3 below.

And pathloss profiles for models calibrated with measurement data for the suburban site are as shown in Fig. (7).

It is important to note that for this case of operating frequency greater than 2GHz, the expression for the multiscreen diffraction component of the basic Walfisch-Ikegami model differs for the COST231 and ITU-R models, in cases where operating frequency is less than 2GHz. In particular, the term '$k_a$', which accounts for how pathloss depends on the relative height of the base station antenna, becomes 71.4, as against 54; and '$k_f$', which determines how multiscreen diffraction loss depends on frequency and distance from transmitter, becomes -8, instead of the expression for '$f_{11}$' in Eqn. (3). It is not surprising therefore, to find that the statistical performance metrics for the case of the 'metropolitan area' differ in pattern from those earlier described for the 900MHz networks

As seen from Table (3), the RMSE and MPE metrics (for the 'metropolitan area) of the calibrated ITU-R (metropolitan) Walfisch-Ikegami differ from corresponding metrics for the other Walfisch-Ikegami models, with that of its calibrated 'suburban' version recording the best metrics of the lot. On the other hand, for the Walfisch-Ikegami models (COST231 and ITU-R) calibrated with measurement data for the suburban site have virtually identical RMSE metrics, with that of the COST231-Metropolitan model being slightly different, in this case. One interesting feature of these computational results is that the calibrated Walfisch-Bertoni model recorded the best RMSE metrics, even better than those for the benchmark ECC33 models.

Contributions to net pathloss by component parameters are described by the profiles of the disaggregated predictions of Fig. (8) and (9) for the QMM-calibrated (using data for metropolitan area of [15]) Walfisch-Ikegami, and Walfisch-Bertoni models, respectively.

The profiles of Fig. (9) follow the same general pattern described earlier for Fig. (3). However, for the profiles of Fig. (9), unlike corresponding profiles of Fig. (5), it is readily observed that contributions to net pathloss concerning 'influence of building geometry' are higher for the basic model than for the QMM-calibrated model. Also, contributions due to the free-space component are in this case, generally higher than corresponding contributions for the calibrated model, whilst the contributions accounting for the curvature of the Earth are about the same, in both cases.

## 4. CONCLUDING REMARKS

This concludes this paper's investigation of the response to Quasi-Moment-Method calibration (QMM) by the COST-231 Walfisch-Ikegami, ITU-R Walfisch-Ikegami, and Walfisch-Bertoni models. First, the paper analytically established that the component parameters of the models satisfy the requirements for use as expansion and testing functions in Moment-Method schemes. And then, using these parameters, the paper developed QMM-calibrated models corresponding to the 'metropolitan' and 'suburban' varieties of the COST231- and ITU-R- Walfisch-Ikegami models, as well as the Walfisch-Bertoni model. Three different sets of measurement data available (through the use of a commercial graph digitizer solution) from the open literature were utilized for the calibration. One from [4], concerning a deployed 900MHz network, for which 'real-time' pathloss measurements were taken; another, also a 900MHz network, with field measurement data (average pathloss measurements) for four different sites, and a third, an LTE (3.4GHz) network, which provided pathloss measurement data for two sites.

Computational results suggest that in general, the calibrated Walfisch-Ikegami models can be expected to respond to QMM-calibration with virtually identical RMSE metrics. The results also indicate that whereas all the 'Walfisch-type' models considered have comparable, and generally excellent MPE responses, the basic Walfisch-Bertoni model, in terms of RMSE, responds better to QMM-calibration than the Walfisch-Ikegami models. Net pathloss predicted by the QMM-calibrated models were disaggregated, in the case of Walfisch-Ikegami models into components associated with free-space, roof-top-to-screen diffraction and scatter loss, multiscreen diffraction; and loss due to effects of building geometry and effects of the Earth's curvature, in the case of the Walfisch-Bertoni models. The contributions due to these components were compared, in a few representative cases, with contributions by components of corresponding basic models.

One notable conclusion arising from these investigations is that in selecting models for pathloss prediction optimization, RMSE performance of the basic model may not represent the best criterion.